\def\rfr#1{eq. (\ref{#1})}

\def\virg#1{``#1''}

\def\eqi{\begin{equation}}
\def\eqf{\end{equation}}
\def\eqia{\begin{eqnarray}}
\def\eqfa{\end{eqnarray}}
\def\rp#1#2{{#1\over#2}} \def\lb#1{\label{#1}}

\documentclass{aastex}
\usepackage{url}
\usepackage{amsmath,amsthm,amscd,amssymb}
\usepackage{latexsym}
\usepackage{graphicx,epsfig}

\RequirePackage{color}

\newcommand{\emaila}{lorenzo.iorio@libero.it}

\begin{document}

\title{Is it plausible to expect a close encounter of the Earth with a yet undiscovered astronomical object in the next few years?}
\shortauthors{L. Iorio}

\author{Lorenzo Iorio\altaffilmark{1} }
\affil{Ministero dell'Istruzione, dell'Universit\`{a} e della Ricerca (M.I.U.R.),\\
Fellow of the Royal Astronomical Society (F.R.A.S.).\\
 Viale Unit\`{a} di Italia 68, 70125, Bari (BA), Italy.}

\email{\emaila}

\begin{abstract}
We analytically and numerically investigate the possibility that a still undiscovered body X, moving along an unbound hyperbolic path from outside the solar system, may penetrate its inner regions in the next few years posing a threat to the Earth.  By conservatively using as initial position the lower bounds on the present-day distance $d_{\rm X}$ of X dynamically inferred from the gravitational perturbations induced by it on the orbital motions of the planets of the solar system, both the analyses show that, in order to reach the Earth's orbit in the next 2 yr, X should move at a highly unrealistic speed $v$, whatever its mass $M_{\rm X}$ is. For example, by assuming for it a solar ($M_{\rm X}=$M$_{\odot}$) or brown dwarf mass ($M_{\rm X}=80m_{\rm Jup}$), now at not less than $d_{\rm X}=11-6$ kau (1 kau=1000 astronomical units), $v$ would be of the order of $6-10\%$ and $3-5\%$  of the speed of light $c$, respectively. By assuming  larger present-day distances for X, on the basis of the lacking of direct observational evidences of electromagnetic origin for it, its speed would be even higher. Instead, the fastest solitary massive objects known so far, like hypervelocity stars (HVSs) and supernova remnants (SRs), travel at $v\approx 0.002-0.005c$, having acquired so huge velocities in some of the most violent astrophysical phenomena like interactions with supermassive galactic black holes and supernova explosions. It turns out that the orbit of the Earth would not be macroscopically altered by a close ($0.2$ au) passage of such an ultrafast body X in the next 2 yr. On the contrary, our planet would be hurled into the space if a Sun-sized body X would encounter it  by moving at $v/c=10^{-4}$. On the other hand, this would imply
 that such a X should be now at just $20-30$ au, contrary to all direct observational and indirect dynamical evidences.
\end{abstract}



\section{Introduction}
Several free-floating astronomical bodies traveling in the interstellar space in the Milky Way have been recently detected.

In recent years  a handful (16) of unbound astrophysical objects  lonely wandering through the Milky Way with speeds as large as about $v\approx 0.1\% c$, where $c$ is the speed of light, have been discovered \citep{Brown05,Edel05,Hir05,Brown06a,Brown06b,Brown07a,Brown07b,Heb08,Brown09,Till09,Brown010,Irr010}. They are the so-called hypervelocity stars (HVSs), whose existence as a consequence of  the Massive Black Hole (MBH) hosted in the center of the Galaxy \citep{Ghez,Gille}, was predicted by \citet{Hills}. Gravitational mechanisms of ejection based on  three-body mutual interactions of binary systems with the MBH, or possibly a pair of MBHs, have been proposed by \citet{Hills} and \citet{Yu}. The consequent rates of HVSs creation would be of the order of $10^{-3}-10^{-4}$ yr$^{-1}$ \citep{Pere07,Yu}. About $10^3$ HVSs may exist within the Galactic solar circle \citep{Yu}. Contrary to those neutron stars exhibiting high proper motions, which are 
supernova remnants (SRs), known HVSs are mostly B-type
main-sequence stars. As an example,  HE $0437-5439$ \citep{Edel05}, moving at heliocentric speed $v=723$ km s$^{-1}=152.517$ au yr$^{-1}=0.0024 c$ \citep{Edel05}, is a B star with mass $M=9$M$_{\odot}$ \citep{Brown010}. The study by \citet{Brown010} has yielded  the first compelling evidence that these HVSs actually come from the center of the Galaxy. All the known HVSs are at about 50 kpc and are unbound with respect to the Galaxy.

Another class of isolated astrophysical objects moving at very fast speeds, not related to  HVSs, is represented by those neutron stars which are the remnants  of asymmetric  explosions of core-collapse supernov{\ae} (SNe) \citep{Sne}. Their extreme speeds are very likely to be attributed to the kick \citep{kick} received in such a kind of peculiar deflagrations\footnote{Indeed,  if the explosion
of a progenitor star expels the ejecta preferentially in one
direction, the compact core must recoil in the opposite direction because of momentum conservation.}.
By measuring the displacements of young pulsars from the apparent
centers of their associated SN shells and using the pulsar
spin-down periods as age estimates, \citet{Cara93} and \citet{Frail94} inferred that pulsars are typically born with transverse
velocities of 500 km s$^{-1}$, and that velocities $v\gtrsim 2000$ km s$^{-1}$ may be possible.
At present, the observational record belongs to the radio-quiet neutron star RX J$0822-4300$, which moves at a record speed of  $1570$ km s$^{-1}=331.191$ au yr$^{-1}=0.0052 c$ at a distance of $7000\ {\rm lyr}=4\times 10^8\ {\rm au}$, as measured in 2007 by the Chandra X-ray Observatory \citep{Wink07}. It is thought to have been produced in an asymmetric SN explosion.

Moving to isolated substellar objects having smaller velocities by about one order of magnitude ($v\approx 10^{-4}c$),
we have the so-called brown dwarfs.
They are astrophysical objects in the range mass $M\approx 0.04-0.09$M$_{\odot}=41-94m_{\rm Jup}$  unable to sustain hydrogen fusion in their cores; as a consequence, it is very difficult to detect them, since most of the energy of gravitational contraction is radiated away within $10^8$ yr, leaving only a very low residual luminosity. After that their existence was postulated for the first time by \citet{brown1} and \citet{brown2}, the first undisputed discovered brown dwarf, and the first T dwarf, was Gl 229B \citep{naka}, with a mass $M=20-50m_{\rm Jup}$. After the advent of large-area surveys
with near-infrared (IR) capability in the late 1990's, hundreds more brown dwarfs were discovered \citep{Kirk}.
Actually, smaller brown dwarf, with $M<20m_{\rm Jup}$, exist \citep{Lod}. In particular, in 2005 \citet{Cha} discovered Cha $110913-773444$. It is a planetary-mass brown dwarf with $M=8m_{\rm Jup}$, which is well  within the mass range observed for bounded extrasolar planets ($M\lesssim 15 m_{\rm Jup}$).
An even smaller body, named rho Oph 4450 with $M=2-3m_{\rm Jup}$, has been recently discovered by \citet{mini}.

Concerning the existence of free-floating  planets of smaller mass, \citet{Steve} noted that, under certain circumstances, Earth-sized solid bodies wandering in the interstellar space after being ejected during the formation of their parent stellar systems may sustain forms of life.
Again as a consequence of three-body interactions with Jovian gas giants, \citet{Debe} have recently shown that during planet formation a non-negligible
fraction of terrestrial-sized planets with lunar-sized companions will likely be ejected into interstellar space with
the companion bound to the planet. \citet{Debe} yield a total number  of free-floating
binary planets in the Galaxy as large as $7\times 10^8$. At present, no planets like them have yet been detected. Proposed microlensing surveys of next generation
will be sensitive to free-floating terrestrial planets \citep{lens}; under certain circumstances, they may be able to yield $10–100$ detections of Earth-mass
free-floating planets \citep{lens}. One to a few detections
could be made with all-sky IR surveys \citep{Debe}.

Are there some solitary traveling astronomical objects, still undetected for some reasons, which may hit the Earth over a time scale of a few years?
 In view of the growing attention that such a possibility  may really occur on\footnote{See, e.g., http://en.wikipedia.org/wiki/Nibiru$\_$collision on the WEB.} 21 December 2012  is receiving in larger portions, also (relatively) educated, of the large public, the present study may also have a somewhat pedagogical/educational value contributing, hopefully, to dissipate certain fears too often artificially induced simply for the sake of gain. Mere academic disdain and/or conceit, derision, and hurling insults should not be retained as adequate practices to counter them. Moreover, the analysis presented here can be repeated in future  when other \virg{doomsday} dates will likely pop out.

The paper is organized as follows. In Section \ref{analisi} we present a relatively simplified analytical calculation\footnote{See also http://www.badastronomy.com/bad/misc/planetx/orbitmath.html on the WEB for the case of a bound, highly eccentric orbit of X coming close to the Earth.} which, however, grasp the essential features of the situation investigated. A more sophisticated numerical analysis is presented in Section \ref{numero}. It is based on the numerical integration of the equations of motion by randomly varying the initial conditions. Section \ref{conclu} summarizes our findings.
 \section{Analytical calculation}\lb{analisi}
 Let us consider a simplified two-body scenario in which a test particle X moves along a heliocentric hyperbola hurling itself towards the Earth. Its conserved (positive) total mechanical energy $E$ is \citep{Landau}
 \eqi E \doteq \rp{1}{2}\mu v^2-\rp{\alpha}{r}>0,\lb{energy}\eqf where $r$ and $v$ are the relative X-Sun distance and speed, respectively, $\mu$ is the system's reduced mass
 \eqi\mu\doteq \rp{M_{\rm X}{\rm M}_{\odot}}{M_{\rm X}+{\rm M}_{\odot}},\eqf and \eqi\alpha\doteq GM_{\rm X}{\rm M}_{\odot},\eqf where $G$ is the Newtonian constant of gravitation.
 The semi-major axis $a$ of the hyperbola is determined by its total energy according to \citep{Landau}
 \eqi a\doteq\rp{\alpha}{2 E}.\eqf
 The eccentricity $e>1$, which, in general, depends on $E$ and on the conserved orbital angular momentum $L$, can be fixed by making the simplifying assumption that the perihelion distance of X \eqi r^{\rm (peri)}\doteq a(e-1)\eqf coincides with, say, the average heliocentric distance of the Earth
 \eqi \left\langle r_{\oplus}\right\rangle = a_{\oplus}\left(1+\rp{e^2_{\oplus}}{2}\right)=1.000142\ {\rm au}.\eqf Thus,
 \eqi e=\rp{\left\langle r_{\oplus}\right\rangle}{a}+1.\eqf
 Thus, $e$ depends now only on the conserved energy $E$.
The parametric equations for the hyperbola are \citep{Landau}
\eqi
\begin{array}{lll}
r & = & a\left(e\cosh\xi -1\right), \\ \\
t &=& \sqrt{\rp{\mu a^3}{\alpha}}\left(e\sinh\xi -\xi\right),
\end{array}\lb{paraorbita}
\eqf
where the parameter $\xi$ takes all values from $-\infty$ to $+\infty$; at perihelion $\xi = 0$.
Let us, now, fix $r$ to a given value. It is the heliocentric distance $d_{\rm X}$ at which the putative X should be located at the present epoch. Remaining in the realm of celestial mechanics, $d_{\rm X}$ can be thought as dynamically constrained by its perturbations of the orbital motions of the known bound major bodies of the solar system. In particular, upper limits on the tidal parameter of X
\eqi \mathcal{K}_{\rm X}\doteq\rp{GM_{\rm X}}{d^3_{\rm X}}\eqf
have been recently obtained \citep{Iorio09} by using the secular precessions of the longitudes of the perihelia $\varpi$ of the inner planets: for each assumed value of the X's mass there is a different lower limit for $d_{\rm X}$. Although, strictly speaking, they have been obtained by assuming X fixed during a planetary orbital revolution, we will use them for the sake of concreteness. Of course, if we quite reasonably postulate that X is made of baryonic\footnote{Concerning the putative existence of stars and planets made of a particular kind of non-baryonic dark matter, i.e. the so-called mirror matter \citep{mirror},  see \citet{Foot1,Foot2,Foot3}.} matter emitting electromagnetic radiation, other, tighter bounds on its present-day distance may be derived from its electromagnetic direct detectability. The recently launched Wide-field Infrared Survey Explorer (WISE)  \citep{Wise} will survey the entire sky in the mid-IR with far greater sensitivity than any previous all-sky IR surveys \citep{surveys} like, e.g., that performed by  the Infrared Astronomical Satellite\footnote{See http://spider.ipac.caltech.edu/staff/tchester/iras/no$\_$tenth$\_$planet$\_$yet.html on the WEB about the alleged discovery of a planet in the remote peripheries of the solar system a by IRAS.} (IRAS) \citep{iras}. Among the scientific goals of WISE there is also the detection of solitary brown dwarf-like bodies in the neighborhood of the solar system. WISE should be able to reveal the existence of a body with the mass of Jupiter within\footnote{See on the WEB http://www.planetary.org/blog/article/00002070/.} $63\ {\rm kau} = 0.3\ {\rm pc}=1\ {\rm lyr}$, while a lightweight brown dwarf with $M_{\rm X}=2-3m_{\rm Jup}$ would be detectable up to $412-618\ {\rm kau} =2-3\ {\rm pc}=7-10\ {\rm lyr}$. Moreover, WISE could find a Neptune-sized object out to 700 au.
Now, by keeping $r=d_{\rm X}$ it is possible to extract the contemporary value of the parameter $\xi$ corresponding to the present-day distance of X
 \eqi\overline{\xi} = {\rm arccosh}\left[\rp{1}{e}\left(1+\rp{d_{\rm X}}{a}\right)\right].\lb{csi}\eqf
 By substituting \rfr{csi} into the parametric equation of $t$ of \rfr{paraorbita}, one can plot the time required to pass from $d_{\rm X}$ to $\left\langle r _{\oplus}\right\rangle$ as a function of the alleged velocity of X at the present epoch: it is sufficient to evaluate $E$ of \rfr{energy} for $r=d_{\rm X}$.  Thus, from the value of the velocity required to take a given time interval-typically of a few yr-to reach the Earth's orbit starting from $d_{\rm X}$, it is possible to make reasonable guesses about the plausibility of the hypothesis that such a putative body X moving towards our planet  actually exists out there.

\begin{figure}[ht!]
 \centerline{\psfig{file=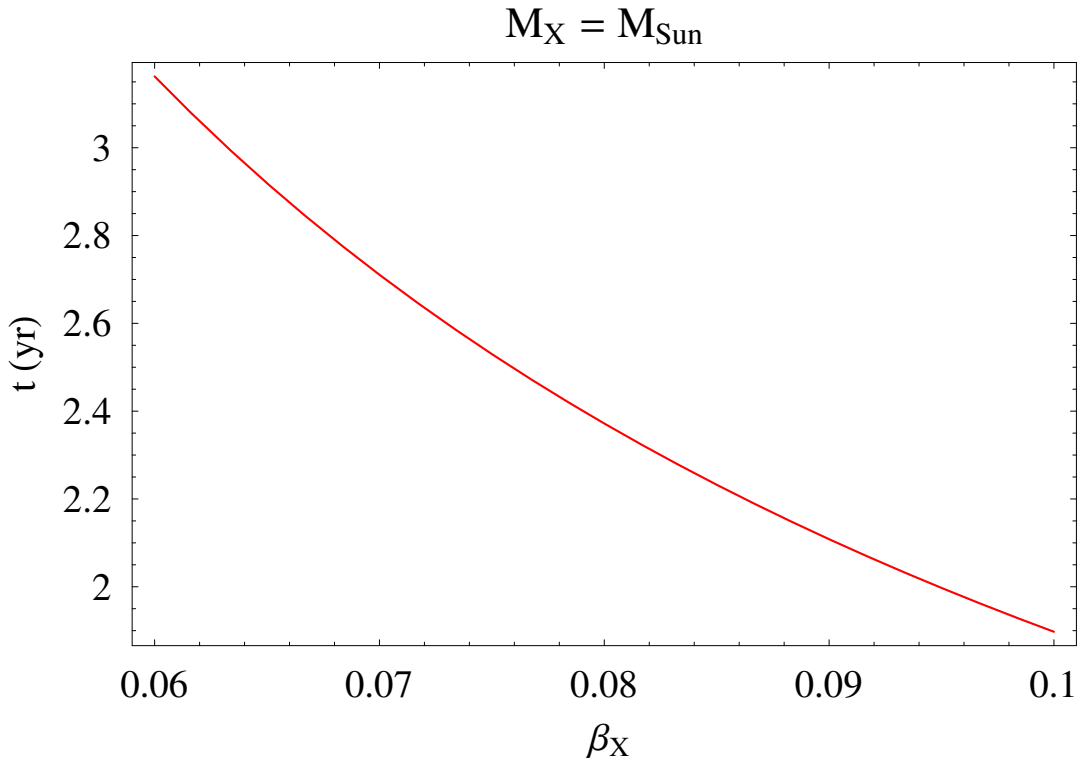,width=\columnwidth}}
\vspace*{8pt}
\caption{Time $t$, in yr, required to a body X with $M_{\rm X}=$M$_{\odot}$ to reach the terrestrial orbit from $d_{\rm X}=12$ kau as a function of its present day speed $\beta_{\rm X}$, in units of $c$.}\label{figura1}
\end{figure}

 To be more specific, let us assume that X is an object with the mass of the Sun; in this case, \citet{Iorio09} yields $d_{\rm X}= 12$ kau as dynamically inferred lower bound of its present heliocentric distance.
 Figure \ref{figura1} shows that such a Sun-sized body X should travel at a implausibly high speed ($v\approx 0.06-0.1 c$) to reach our orbit in a few years from now.
 Recall that the highest recorded speeds of unbound objects of stellar size are $0.002-0.005c$. Note that the situation is even worse if we take a larger value for the limit distance $d_{\rm X}$ in view of the fact that, after all, a baryonic star should have been easily detected if it was really at just 12 kau from us. Indeed, it turns out that by setting, say, $d_{\rm X}=100$ kau the required speed would closely approach $c$. It may be of interest to note that, by traveling at $v=0.002-0.005c$, a Sun-sized body X would take $300-800$ yr to reach our orbit if it was now at 100 kau from us, while $40-90$ yr would be required if it was at just 12 kau. Incidentally, let us remark that the closest black hole so far discovered, whose distance has been directly measured from its parallax using astrometric VLBI observations,  is in the X-ray binary V404 Cyg, at about $2\ {\rm kpc}=4\times 10^8\ {\rm au}$  \citep{BH2}. Another close black hole is V4641 Sgr \citep{BH1}, at about $7-12\ {\rm kpc}$.

\begin{figure}[ht!]
 \centerline{\psfig{file=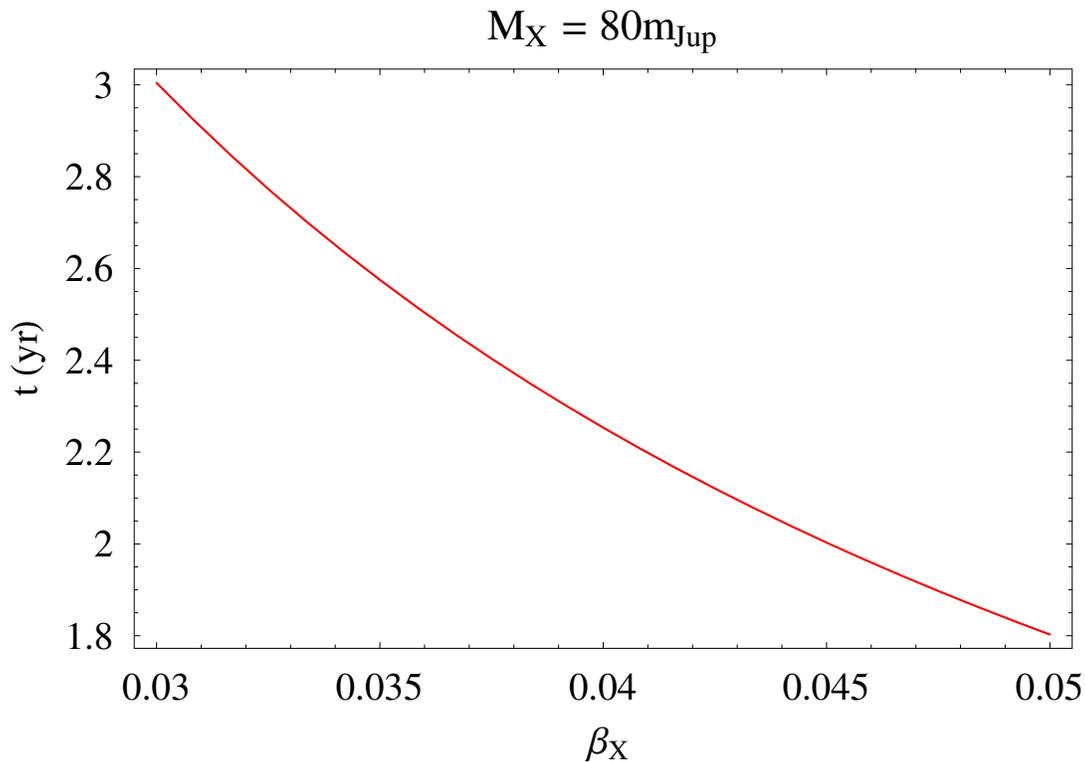,width=\columnwidth}}
\vspace*{8pt}
\caption{Time $t$, in yr, required to a body X with $M_{\rm X}=80m_{\rm Jup}$ to reach the terrestrial orbit from $d_{\rm X}=5.7$ kau as a function of its present day speed $\beta_{\rm X}$, in units of $c$.}\label{figura2}
\end{figure}

 Figure \ref{figura2} depicts the case of a brown dwarf with $M=80m_{\rm Jup}$ and $d_{\rm X}= 5.2$ kau \citep{Iorio09}.
Also such a scenario looks highly implausible because it should be $v/c\approx 0.03-0.05$; moreover, for $d_{\rm X}=20$ kau it turns out that $v/c\approx 0.08-0.2$.
No brown dwarfs at all  moving at speeds comparable to those of SNRs and HVSs are known; on the contrary, their speeds are of the order of $v\approx 100$ km s$^{-1}=3\times 10^{-4}c$ \citep{Faherty}. Traveling at such typical speeds, it would take $1-3$ kyr to reach the terrestrial orbit for $d_{\rm X}=20$ kau, and $300-900$ yr for $d_{\rm X}=5.7$ kau.

\begin{figure}[ht!]
 \centerline{\psfig{file=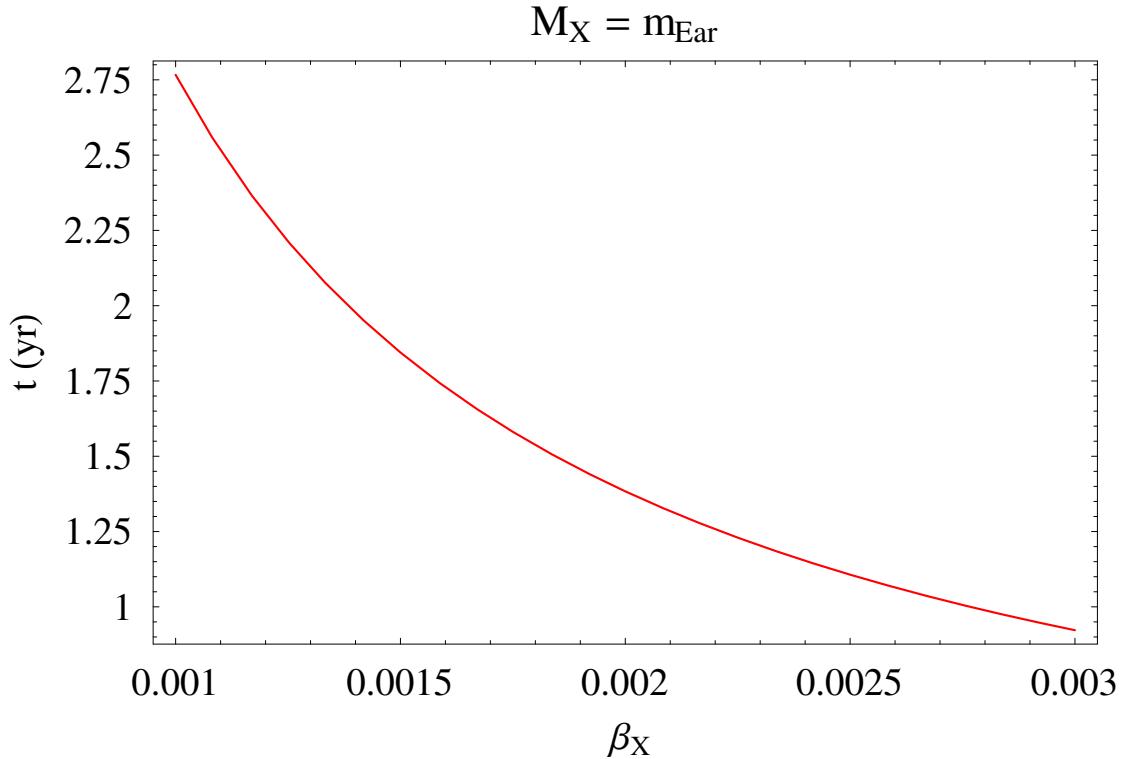,width=\columnwidth}}
\vspace*{8pt}
\caption{Time $t$, in yr, required to a body X with $M_{\rm X}=m_{\oplus}$ to reach the terrestrial orbit from $d_{\rm X}=175$ au as a function of its present day speed $\beta_{\rm X}$, in units of $c$.}\label{figura3}
\end{figure}

Figure \ref{figura3} shows that also the case in which the putative colliding X is a rock-ice body with the mass of the Earth is unlikely because, by assuming $d_{\rm X}=175$ au \citep{Iorio09}, it should travel at $v/c=0.001-0.003$ to reach the orbit of our planet in the next few yr.
Given the ejection mechanisms occurring in the planet formation processes which may be responsible for such free-floating small planets, their typical velocities should be of the order of $v\approx 1-3$ km s$^{-1}=1-0.3\times 10^{-5} c$ for a Jupiter-sized mass ejecting body \citep{Gold}. Thus,  $180-300$ yr
would be required by traveling at such speeds if an Earth-sized body X was now at $d_{\rm X}=175$ au.

 As we will see in Section \ref{numero}, the conclusions of such a simplistic analytical two-body scenario are also supported by a more sophisticated, numerical analysis.

 It may be interesting to note that some reflections by M. Brown similar to the reasonings developed in detail in this Section can be found at http://news.discovery.com/space/mike-brown-planetx-pluto.html on the Internet. The case of a body, of unspecified mass, reaching the Earth's orbit on an unbound trajectory in the next 2 yr starting  now from 1 kau  is touched. Strictly speaking, the speed of such an unbound X is  computed by assuming that it travels uniformly, so that it is $v=2.4\times 10^3$ km s$^{-1}=0.008c$. According to \citet{Iorio09}, 1 kau is the dynamically inferred lower limit for a body with $M_{\rm X}=m_{\rm Jup}$ lying perpendicularly to the ecliptic; the speed required to come here in the next $2-1.6$ yr turns out to be $0.8-1\%$ of $c$. If we take $d_{\rm X}=1.2$ kau for a jovian-sized body lurking now in the ecliptic \citep{Iorio09}, we get $v/c=0.01$ to reach 1 au in the next 2 yr.
 Concerning a Jupiter-sized body X, Brown at http://news.discovery.com/space/mike-brown-planetx-pluto.html puts it at at a few thousand au; in this case, by setting, say, $d_{\rm X}=2.5$ kau we have $v/c=0.02$.
\section{Numerical calculation}\lb{numero}
We, first, numerically integrated the equations of motion  of an unbound body X in a ICRF/J2000.0 heliocentric frame with a coordinate system employing rectangular Cartesian coordinates along with the ecliptic and mean equinox of reference epoch J2000. In regard to the  initial position chosen for X, we took the predicted coordinates of the Earth at $t_0=$21 December 2012 retrieved from the HORIZONS WEB interface by NASA/JPL and added randomly generated small corrections to them, i.e.,
\eqi
\begin{array}{lll}
x_0^{\rm (X)} &=& x_{\oplus}(t_0)+\delta_x,\\ \\
y_0^{\rm (X)} &=& y_{\oplus}(t_0)+\delta_y,\\ \\
z_0^{\rm (X)} &=& z_{\oplus}(t_0)+\delta_z,\\ \\
\end{array}
\eqf
where $\delta_x,\delta_y,\delta_z$ were randomly generated from a uniform distribution within $\pm 0.001$ au.
Concerning the initial velocity, we randomly generated it by imposing the conditions
\eqi
\begin{array}{lll}
v_0^{\rm (X)} & > & \sqrt{\rp{2G(M_{\odot}+M_{\rm X})}{r_0^{\rm X}}},\\ \\
v_0^{\rm (X)} & < & c,
\end{array}
\eqf
where
\eqi v_{p} \doteq \sqrt{\rp{2G({\rm M}_{\odot}+M_{\rm X})}{r^{\rm X}}}\eqf is the limit parabolic velocity; a hyperbola occurs if $v>v_{p}$.
Starting from such sets of randomly generated initial conditions, we numerically propagated the trajectory of X backward in time over 2 yr, so that  $t_{\rm fin}$ represents the present-day epoch.
In such a way, by performing several runs, the conclusions of  Section \ref{analisi} turned out to be substantially confirmed in the sense that, in order to avoid  finding X at the end of the integration, i.e. at the present epoch, closer than the dynamically inferred lower limits $d_{\rm X}$, too high velocities would be required.

\begin{figure}[ht!]
 \centerline{\psfig{file=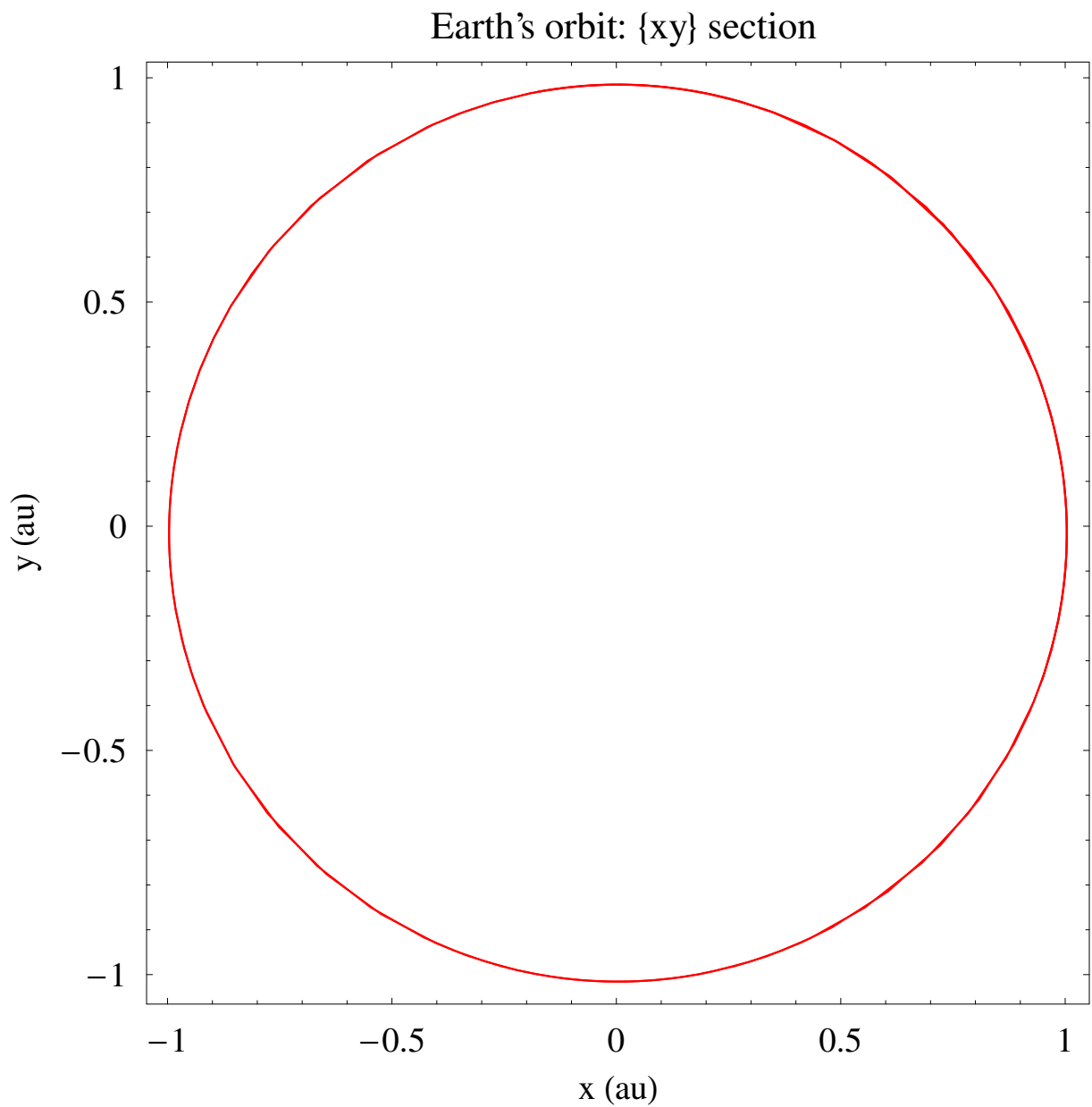,width=\columnwidth}}
\vspace*{8pt}
\caption{Section in the $\{xy\}$ plane of the numerically integrated Earth's orbit over the next 2 yr by assuming for X $M_{\rm X}=$M$_{\odot}$, $d_{\rm X}=11.241$ kau, $v=0.076c$.
}\label{figura4}
\end{figure}

\begin{figure}[ht!]
 \centerline{\psfig{file=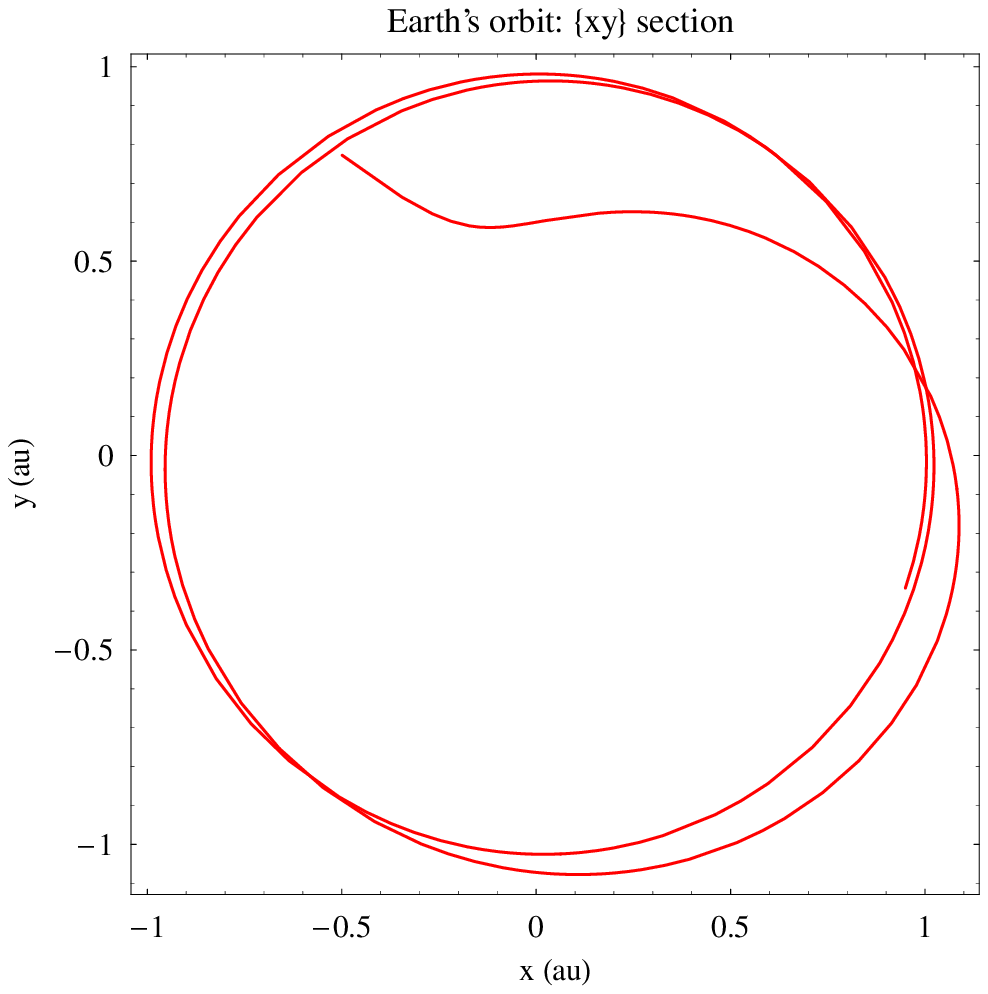,width=\columnwidth}}
\vspace*{8pt}
\caption{Section in the $\{xy\}$ plane of the numerically integrated Earth's orbit over the next 2 yr by assuming for X $M_{\rm X}=$M$_{\odot}$, $d_{\rm X}=26.3$ au, $v=2.2\times 10^{-4}c$.
}\label{figura5}
\end{figure}

Then, we made a further numerical analysis in which we used the final state vectors of X of each of the previous runs backward in time as initial conditions for new runs performed, now, forward in time over 2 yr. In other words, now $t_0$ corresponds to the present epoch, while $t_{\rm fin}=21\ {\rm December}\ 2012$. In such new runs we also added the Earth, Jupiter and Saturn by modeling their mutual interactions and their attractions on X. Their initial conditions, corresponding to the present epoch, were retrieved from the HORIZONS WEB interface. The situation remains unaltered: starting today from positions corresponding to the dynamically inferred lower limits $d_{\rm X}$, all the numerically propagated trajectories of X reach  heliocentric distances of about 1 au in next 2 yr traveling at unrealistically high speeds, as seen in Section \ref{analisi}. It turns out that, also according to such an analysis, larger initial distances for X yield even larger speeds for it, just as in Section \ref{analisi}. The inclusion of the major planets of the solar system do not cause noticeable alterations to such a picture. Conversely, from our numerical analysis it turns out that the hypothetical passage of such a fast body X would not distort the orbits of the planets considered, in particular of the Earth. This is clearly depicted by Figure \ref{figura4} which shows the numerically integrated terrestrial orbit in the next 2 yr in the case of a Sun-sized X body supposed located today at $11.241$ kau and moving with $v/c=0.076$.
Incidentally, in the example showed the mutual X-Earth distance at $t_{\rm fin}$ amounts to $\Delta r=0.2$ au. Instead, a much smaller velocity of
X would induce macroscopically noticeable changes in the Earth's orbit, as shown in Figure \ref{figura5}. It is obtained for $v/c=2.2\times 10^{-4}$, with $v/v_{p}=1.1$.
Such a scenario would be catastrophic since in it the Earth would be finally stripped from its orbit and thrown away, as an extension of the time span of the numerical integration to 5 yr shows. Of course, it is highly unrealistic since it  implies the present existence of an undiscovered Sun-sized body X  at just 26.3 au.
\section{Summary and conclusions}\lb{conclu}
We analytically and numerically investigated the possibility-which cyclically gains popularity for a variety of psychological and/or sociological reasons in extended portions of the large public, even cultivated-that a yet undiscovered astronomical body X, moving on an unbound trajectory from outside the solar system, may penetrate its inner regions by  closely encountering the Earth in the next few years. For the sake of concreteness we choose a time span of 2 yr ending at 21 December 2012, familiar to  a non-negligible amount of people, but the strategy outlined here can naturally be extended to any temporal interval and dates in the not unlikely case that in the more or less near future-presumably  after 2012-other analogous \virg{doomsdays}  of astronomical origin will be proposed.

As initial positions, we conservatively choose the lower limits $d_{\rm X}$ for the present-day distance of such a putative  X from the bounds dynamically inferred from the magnitude of the perturbations that it would induce on the orbital motions of the inner planets of the solar system. Given that, at present, there are no direct observational evidences of electromagnetic origin for the existence of X, tighter constraints on its distance, i.e. larger values for $d_{\rm X}$, may  well have been adopted. The initial velocities were chosen by allowing for unbound, hyperbolic trajectories in the field of the Sun. Both analytical and numerical calculations, performed for different values of the mass of X by randomly varying its initial conditions and including also the Earth, Jupiter and Saturn, show that, in all cases, X should move at unrealistically high velocities to reach heliocentric distances of 1 au in the next 2 yr. No known astrophysical objects with  high speeds, acquired in certain known physical processes, move as fast as the putative X should do. In the case of a body with the mass of the Sun or of a typical brown dwarf ($M_{\rm X}=80 m_{\rm Jup}$) the speed required to come close the Earth in the next 2 yr from presently assumed distances of thousands-ten thousands astronomical units would be $6-10\%$ and $3-5\%$ of  $c$, respectively. Even higher speeds are involved if we adopt larger values of the initial distance of X relying upon the still missing direct detection of it from electromagnetic radiation. The fastest Sun-sized objects known so far travel at speeds as large as $0.2-0.5\%$  of $c$, and are produced in some of the most violent astrophysical processes known  like interactions with supermassive galactic black holes and supernova deflagrations.
Moreover, it turns out that the orbit of the Earth would not be distorted in a macroscopically noticeable way by the close ($0.2$ au) passage of such a hypothetical ultrafast body. The terrestrial path would be sensibly altered in such a way that the Earth would be thrown away  if the speeds involved by a passing star-sized body were quite smaller, of the order of $0.01\%$ of $c$. Such a scenario is highly unrealistic because, in this case, X should be now at just a few ten astronomical units.
%


\end{document}